\documentclass[useAMS,usenatbib,usegraphicx]{mn2e}
\onecolumn{}

\usepackage[utf8]{inputenc}

\title[WASP 1625-04]{The Pre-He White Dwarfs in Eclipsing Binaries. III. WASP 1625-04}

\author[J. W. Lee et al.]
       {Jae Woo Lee$^{1}$\thanks{E-mail: jwlee@kasi.re.kr}, Kyeongsoo Hong$^{2}$ and Jang-Ho Park$^{1}$ \\
        $^1$Korea Astronomy and Space Science Institute, Daejeon 34055, Republic of Korea \\
        $^2$Institute for Astrophysics, Chungbuk National University, Cheongju 28644, Republic of Korea }

\begin{document}

\date{Accepted 2021 ---------. Received 2021 ---------; in original form 2021 }

\pagerange{\pageref{firstpage}--\pageref{lastpage}} \pubyear{2021}

\maketitle

\label{firstpage}

\begin{abstract}
1SWASP J162545.15-043027.9 (WASP 1625-04) has been announced as one of EL CVn candidates showing total primary eclipses and 
ellipsoidal variations. This paper presents the absolute properties of the binary star, based on our high-resolution spectroscopy 
conducted from 2015 through 2020. From the spectral analysis, the radial velocities (RVs) for both components were obtained 
with the effective temperature $T_{\rm eff,1} = 8990 \pm 200$ K and the rotational rate $v_1\sin$$i=53\pm5$ km s$^{-1}$ 
for the more massive primary. The RV measurements were analyzed with archival WASP photometry. From the modelling we obtained: 
$M_1 = 1.745 \pm 0.013$ M$_\odot$, $M_2 = 0.187 \pm 0.002$ M$_\odot$, $R_1 = 1.626 \pm 0.008$ M$_\odot$, $R_2 = 0.290 \pm 0.003$ M$_\odot$, 
$L_1 = 15.5 \pm 1.4$ L$_\odot$, and $L_2 = 1.84 \pm 0.16$ L$_\odot$. In the Hertzsprung-Russell diagram, WASP 1625-04 A lies on 
the zero-age main sequence and its companion accords well with the helium-core white dwarf models of 0.19 M$_\odot$ in the constant 
luminosity phase. Our improved results demonstrate that WASP 1625-04 is a typical EL CVn-type binary with a low mass ratio 
and $M_2$ combination in the thin-disk population and is the product of the stable, non-conservative mass transfer of the precursor binary. 
\end{abstract}


\section{INTRODUCTION}

Double-lined eclipsing binaries (EBs) provide us with robust determination of fundamental stellar parameters. 
With time-series spectroscopic and photometric data, we can measure the masses and radii of each component to precisions 
of 3 \% or better (Torres, Andersen \& Gim\'enez 2010; Maxted et al. 2020). Such measurements have been used to evaluate 
various theoretical models and to improve them. The best-studied EBs are therefore prime objects for understanding 
stellar structure and evolutionary processes. Of particular interest are EBs with helium-core white dwarfs (WDs) of 
$M_2 \la$ 0.3 M$_\odot$ as they cannot evolve from single stars (Marsh, Dhillon \& Duck 1995; Kilic, Stanek \& Pinsonneault 2007). 
This is our third paper in a series detecting low-mass WDs in EB systems and studying their physical properties, 
from new high-resolution spectra combined with photometric data (Lee et al. 2020, Paper I; Hong et al. 2021, Paper II). 

EL CVn-type stars are post-mass transfer EBs in which the primary component is an A/F-type main sequence and the secondary one 
is a He-core WD precursor (pre-He WD) (Maxted et al. 2011, 2014; van Roestel et al. 2018). These star systems may be 
the product of non-conservative binary evolution (Chen et al. 2017), where the initially primary star has lost considerable mass 
during mass transfer to the red giant branch (RGB), becoming the present pre-He WD secondary, while the mass gainer became 
the present brighter component as the result of mass accretion. Then, the extremely low-mass WDs are the stellar remnants of 
the more massive primary component in the precursor binary system, when there is not enough mass to burn the He in its core. 
Among short-period Algol-type EBs, R CMa stars (Budding \& Butland 2011; Lee et al. 2016) share the characteristics of 
both low mass ratio $q$ and secondary's mass $M_2$ with the EL CVn stars. They are divided into semi-detached and 
detached R CMa types. The semi-detached systems stop their mass transfer, contract, and then evolve into the detached states. 
The detached R CMa stars are thought to be just-born EL CVn stars (Lee et al. 2018). 

In this study, we present 1SWASP J162545.15-043027.9 (WASP 1625-04; HD 148070; TYC 5043-1022-1; Gaia EDR3 4355789947085032448). 
The binary star was discovered by Maxted et al. (2014) to be an EL CVn candidate with an eclipsing period of 1.5263234 days, 
using the photometric data from the WASP survey. From their own spectra, a systemic velocity of $\gamma$ = 9 $\pm$ 4 km s$^{-1}$ 
was obtained and the spectral type of the brighter component (WASP 1625-04 A) was assigned as A2 V. They further analyzed 
the observed flux distribution and the WASP archive data, and obtained binary model parameters, as follows: effective temperatures 
$T_1$ = 9500 $\pm$ 500 K and $T_2$ = 11,500 $\pm$ 1500 K, an orbital inclination of $i$ = 82.5 $\pm$0.3 deg, a mass ratio of 
$q$ = 0.126 $\pm$ 0.026, and relative radii $r_1$ = 0.2381 $\pm$ 0.0025 and $r_2$ = 0.0405 $\pm$ 0.0005. The secondary component 
(WASP 1625-04 B) was expected to be a pre-He WD with a mass of $<$ 0.3 M$_\odot$ in a thin disk Galactic population. 

The rest of the paper consists of the following. Section 2 describes the WASP public data and our spectroscopic observations. 
The double-lined radial velocities (RVs) and the primary star's atmospheric parameters are measured in Section 3. In Section 4, 
we present the fundamental stellar parameters from updated binary modeling, satisfying the RV and light curves. Finally, 
Section 5 discusses the evolutionary history of the target star with our conclusions.

\section{OBSERVATIONS AND DATA REDUCTIONS}

\subsection{WASP PHOTOMETRY}

WASP 1625-04 was observed as part of the Wide Angle Search for Planets (WASP) survey to find transiting exoplanets 
(Pollacco et al. 2006; Wilson et al. 2008). The WASP archive data (Butters et al. 2010) of our target star were provided to us 
by Prof. Pierre Maxted. A total of 54,263 observations were obtained between August 2007 and March 2014 using three cameras: 
166 and 33,611 points, respectively, from cameras 142 and 146 (WASP-North on La Palma, Spain), and 20,486 from camera 
224 (WASP-South at SAAO, South Africa). They had many outliers showing unrealistically large deviations from the mean trend 
of the light curve due to some issues, including weather conditions or instrumental effects. In order to exclude them from 
our analysis, we formed a mean light curve in phase intervals of 0.001 and repeatedly removed data points larger than 
3-$\sigma$ values from it. Finally, we selected 50,184 individual points (30,909 from camera 146 and 19,275 from camera 224) 
corrected using the SYSREM detrending algorithm (Tamuz, Mazeh \& Zucker 2005) in this paper. In Figure 1, the WASP light curve 
of the program target is depicted as magnitudes versus orbital phases.

\subsection{NEW SPECTROSCOPY}

The spectroscopic observations for WASP 1625-04 were conducted during eight nights between 2015 April and 2020 April, using 
the 1.8 m telescope and the BOES echelle spectrograph at the Bohyunsan Optical Astronomy Observatory (BOAO) in Korea. 
The observations covered a wide wavelength range from 3600 to 10,200 $\rm \AA$ with a spectral resolving power ($R$) of 30,000. 
More information about the instrument was detailed in Kim et al. (2007). The total 48 high-resolution spectra were 
obtained with an integration time of 1800 sec and reduced by utilizing the CCDRED and ECHELLE packages in IRAF (Hong et al. 2015). 
The exposure time corresponds to 0.014 of the orbital period, and therefore orbital smearing can be neglected in our analysis. 
The signal-to-noise (S/N) ratios were extracted using the `m' keystroke of the IRAF $splot$ task in the continuum regions 
between 4000 and 5000 $\rm \AA$. Their typical values were estimated as about 35$-$40.

\section{SPECTRAL ANALYSIS} 

In order to measure the double-lined RVs of WASP 1625-04, we followed the same method as that in Hong et al. (2019). First 
of all, we searched for sets of absorption lines from WASP 1625-04 A and B at full wavelength, and found the isolated lines that 
clearly identify the two in only Mg II region. In Figure 2, the phase-folded trailed spectra of Mg II $\lambda$4481 display 
the sinusoidal tracks caused by the orbital motions of each component. The absorption region was also used for the RV measurements 
of the EL CVn systems WASP 0131+28 (Lee et al. 2020) and WASP 0843-11 (Hong et al. 2021). We fitted the Mg II lines with 
double Gaussian functions of the IRAF $splot$ task. As a consequence, the primary and secondary RVs were derived for first time, 
and are presented in Table 1 and Figure 3. In this table, some $V_2$ values are missing because they could not be measured by 
the low S/N ratio and faintness of their observed spectra. We assumed a circular orbit ($e$ = 0) for the EB system, based on 
the phase difference between two minima in the WASP light curve (Figure 1) and the sine wave of our RV curves (Figure 3). 
Without considering the proximity effects such as ellipticity, reflection, and rotation, the circular orbital elements were 
obtained by separately applying each component's RVs to a sine curve (Lee et al. 2018). The sine-curve solution is given in Table 2, 
which lists the reference epoch $T_{\rm 0}$ and period $P$ of the orbital ephemeris, the systemic velocity $\gamma$, 
the RV semi-amplitude $K$, the derived semimajor axis $a\sin$$i$, mass $M\sin$$i$, and mass ratio $q$, and the rms residuals 
from the sine-curve fit. $P$ = 1.5263234 days was taken from Maxted et al. (2014). 

The atmospheric parameters of WASP 1625-04 were yielded by using the $\chi^{2}$ minimization method between disentangling spectra
and theoretical stellar models. For this process, the disentangling spectra of WASP 1625-04 A and B were constructed with 
the FDB\textsc{inary} code, which is based on a Fourier decomposition technique (Iliji\'c et al. 2004). Their S/N ratios were measured 
to be about 100 and 60 in the same order as before. However, as shown in Figure 4, we could not obtain a high-quality spectrum for 
the secondary star, because of its low contribution $L_{\rm B}$/$L_{\rm A}$ = 6 \% (Maxted et al. 2014). 
From the observed flux distribution and their low-resolution spectra, Maxted et al. (2014) estimated the effective temperature and 
spectral type of the primary star to be $9500\pm500$ K and A2 V, respectively. Thus, our theoretical models were calculated using 
surface temperatures ($T_{\rm eff,1}$) of 8000 K$-$12,000 K with steps of 10 K and rotational velocities ($v_1\sin$$i$) of 
10$-$200 km s$^{-1}$ with steps of 1 km s$^{-1}$ based on the BOSZ spectral library\footnote{https://archive.stsci.edu/prepds/bosz} 
(Bohlin et al. 2017). In these grids, the solar metallicity was applied and the surface gravity was kept fixed to $\log$ $g$ = 4.3 
(cf. Section 4). The model spectra were converted from vacuum to air wavelengths using the IAU standard given in Morton (1991). 

Finding the optimal values for $T_{\rm eff,1}$ and $v_1\sin$$i$, we selected five spectral regions, such as Ca II K $\lambda$3933, 
Fe I $\lambda$4046, H$_{\rm \gamma}$, Mg II $\lambda$4481, and H$_{\rm \beta}$, which are useful in temperature classifications 
(Gray \& Corbally 2009). As a result of this process, we obtained the two atmospheric parameters of $T_{\rm eff,1}=8990 \pm 200$ K and 
$v_1\sin$$i=53\pm5$ km s$^{-1}$. These values and their uncertainties are the mean and standard deviation of the best-fit values obtained 
in these regions. The disentangling and synthetic spectra for the chosen spectral regions are presented as circles and red lines in 
Figure 4, respectively.

\section{BINARY MODELING}

Despite its rather large scatter, the light curve of WASP 1625-04 in Figure 1 indicates that the primary minimum is a total eclipse. 
As expected for EL CVn type (Maxted et al. 2014), this feature may be ascribed to the occultation of the pre-He WD secondary 
(WASP 0131+28 B) by the more massive but cooler primary (WASP 0131+28 A). Maxted et al. (2014) reported a photometric solution 
of the system analyzing the WASP light curves with the JKTEBOP code (Southworth 2010). For the binary star model, we used 
our RV curves and the selected WASP data, and applied the detached mode 2 of the Wilson-Devinney (W-D) modeling code 
(Wilson \& Devinney 1971, van Hamme \& Wilson 2007) to the two kinds of observations. The proximity corrections of the RVs were also 
considered. The modeling process is almost identical to the method performed for OO Dra (Lee et al. 2018) and WASP 0131+28 (Lee et al. 2020). 
Here, the subscripts 1 and 2 denote WASP 1625-04 A and B, respectively. 

Based on our spectroscopic results, the effective temperature of WASP 1625-04 A was set to be $T_1$ = 8990 $\pm$ 200 K, 
and the orbital parameters of $a$, $\gamma$, and $ q$ were initialized from Table 2. In the same way as Lee et al. (2020), we used 
the bolometric albedos of $A_{1,2} = 1.0$, the gravity-darkening exponents of $g_{1,2} = 1.0$, and the logarithmic limb-darkening law 
with coefficient interpolated from the tables of van Hamme (1993). The light and RV curves were repeatedly analyzed until both of 
them were satisfied. In this modeling, ten parameters are adjustable: the eclipse ephemeris ($T_0$, $P$), the spectroscopic elements 
($a$, $\gamma$, $q$), the inclination angle ($i$), the secondary's temperature ($T_2$), the surface potentials ($\Omega_{1,2}$), 
and the primary's luminosity ($l_1$). The final solution is illustrated in Table 3 and shown as the solid lines in Figures 1 and 3. 
The corresponding residuals are presented in the lower panels of each figure. Here, they indicate that our binary modeling is 
a good solution of the two datasets. At this point, the orbital eccentricity and argument of periastron were added to 
the fitted parameters. The search retained a value of zero ($e=0$) within a margin of error. 
Recently, Lagos et al. (2021) detected tertiary components in all of the five EL CVn binaries that they observed. The result 
implies that most of the EL CVn stars are at least triplets. Thus, we examined whether a third light source existed or not in 
the WASP public data, but none was found, with an upper limit of 0.001. 

From the combined solution of the RV and light curves in Table 3, we determined straightforwardly the absolute parameters for 
WASP 1625-04. For this computation, we adopted the solar values of $T_{\rm eff}$$_\odot$ = 5780 K and $M_{\rm bol}$$_\odot$ = 
+4.73 and the empirical calibration of $\log T_{\rm eff}$ and bolometric corrections (BCs) in Torres (2010). The results are shown 
in Table 4. The masses and radii were measured to 0.7 \% and 0.5 \% precision for the massive primary component, respectively, and 
1.1 \% and 1.0 \% for the hotter secondary. For the distance determination to WASP 1625-04, we used its apparent magnitude and 
color excess at $V$ = +10.362 $\pm$ 0.029 and $E$($B-V$) = +0.158 $\pm$ 0.004, respectively, extracted from the APASS $BVg'r'i'$ 
photometry (Munari et al. 2014). In consequence, our distance of 420 $\pm$ 22 pc is concurrent with the GAIA EDR3 measurement 
of 408 $\pm$ 6 pc taken from a parallax of 2.449 $\pm$ 0.033 mas (Gaia Collaboration et al. 2021).

\section{DISCUSSION AND CONCLUSIONS}

We reported and studied our echelle spectroscopy of the eclipsing binary star WASP 1625-04, together with the WASP public data. 
The RV measurements of both components were obtained from the BOES spectra. Utilizing the $\chi^{2}$ minimization method, we found 
the primary's temperature to be $T_{\rm eff,1}=8990 \pm 200$ K and the projected rotational rate to be $v_1\sin$$i=53\pm5$ km s$^{-1}$. 
A simultaneous modeling of the RV and WASP data, combined with the atmospheric parameters, robustly yielded the fundamental parameters 
for our target star. The mass and radius of WASP 1625-04 A were determined to a precision of better than 0.7 \% and those 
of its companion to about 1 \%. Our modeling results represent that WASP 1625-04 is a short-period detached binary with both low $q$ 
and $M_2$, typical of the EL CVn type. Calculated from each star's radius and the binary period, the synchronous rotations are 
$v_{\rm 1,sync}$$\sin$$i$ = 53.9 $\pm$ 0.3 km s$^{-1}$ and $v_{\rm 2,sync}$$\sin$$i$ = 9.6 $\pm$ 0.1 km s$^{-1}$, respectively, 
wherein the former value is in line with our BOES measurement.

Using the solar motion of ($U_\odot$, $V_\odot$, $W_\odot$) = (9.58, 10.52, 7.01) km s$^{-1}$ (Tian et al. 2015) in the local standard 
of rest, we calculated the space velocity of WASP 1625-04 to be ($U$, $V$, $W$)\footnote{Positive toward the Galactic center, 
Galactic rotation, and North Galactic Pole.} = (29.0$\pm$0.1, 241.4$\pm$0.4, 11.6$\pm$0.3) km s$^{-1}$ using the formulas 
given by Johnson \& Soderblom (1987). Here, the proper motion ($\mu _\alpha \cos \delta$, $\mu _ \delta$) is taken from the GAIA EDR3, 
and the system velocity ($\gamma$) and distance are from our results. Then, using the ORBIT6 code of Odenkirchen \& Brosche (1992), 
we computed the $z$-component of the angular momentum and the eccentricity of the system's Galactic orbit to be 
$J_{\rm z}$ = 1847 $\pm$ 92 kpc km s$^{-1}$ and $e$ = 0.1320 $\pm$ 0.0004, respectively. For more details about this procedure see 
Lee et al. (2020). The kinematic data of WASP 1625-04 agrees with the thin-disk population in the $U-V$ and $J_{\rm z}-e$ diagrams 
of Pauli et al. (2006). 

To understand the evolutionary history of WASP 1625-04, the positions of both components in the Hertzsprung-Russell (HR) diagram 
are displayed in the left panel of Figure 5, where the blue, green, and cyan lines are the He-core WD models with metallicities 
$Z$ of 0.001, 0.01, and 0.02, respectively, for a mass of 0.19 M$_\odot$ (Istrate et al. 2016). In the HR diagram, 
the more massive primary component is located on the zero-age main sequence (ZAMS), while the hotter companion with 
0.187 $\pm$ 0.002 M$_\odot$ is in the constant luminosity phase and consistent with both models of $Z$ = 0.001 and 0.01. This 
clearly shows that WASP 1625-04 B is a pre-He WD. From the $M_{\rm WD}-t$ relations presented by Chen et al. (2017), we calculated 
the lifetime ($t$) during the pre-He WD stage of WASP 1625-04 B to be about 7.4 $\times$ 10$^{8}$ yr. 

Meanwhile, it is known that extremely low-mass WDs less than $\sim$0.3 M$_\odot$ are generally formed by two channels: 
common-envelope ejection and stable Roche-lobe overflow (cf. Li et al. 2019). Recently, Chen et al. (2017) argued that 
the latter channel is the most likely formation process of the EL CVn binaries. In the case of stable mass transfer, 
Lin et al. (2011) reported that there is a tight relation between the WD masses $M_{\rm WD}$ and the binary periods $P$. 
We put WASP 1625-04 B on the $M_{\rm WD}-\log P$ diagram in the right panel of Figure 5, together with 
eight double-lined R CMa (Wang et al. 2019) and five EL CVn stars: KOI-81 (Matson et al. 2015), WASP 0131+28 (Lee et al. 2020), 
WASP 0247-25 (Kim et al. 2021), WASP 0843-11 (Hong et al. 2021), and WASP 1323+43 (EL CVn, Wang et al. 2020). As one can see in 
this panel, almost all the stars except KOI-81, which has a relatively long period of $P$ = 23.8776 days, seem to coincide with 
the theoretical relation of Lin et al. (2011). The results indicate that most of the pre-He WD companions in the R CMa and 
EL CVn binaries are the product of the stable mass transfer episode, and that WASP 1625-04 is a post-mass transfer binary system, 
comprising an A2 V gainer and a pre-He WD of mass 0.187 M$_\odot$.

\section*{Acknowledgments}
This paper was based on observations obtained at the BOAO, which is operated by the Korea Astronomy and Space Science Institute 
(KASI). We would like to thank the BOAO staffs for assistance during our spectroscopy and Prof. Pierre Maxted for sending us 
the WASP archival data. Funding for WASP comes from the consortium universities and from the UK's Science and Technology 
Facilities Council. We appreciate the careful reading and valuable comments of the anonymous referee. 
This research was supported by the KASI grant 2021-1-830-08. K.H. was supported by the grants 2019R1A2C2085965 and 2020R1A4A2002885 
from the National Research Foundation (NRF) of Korea.

\section*{DATA AVAILABILITY}
The data underlying this article will be shared on reasonable request to the first author.

\clearpage
\begin{figure}
\includegraphics{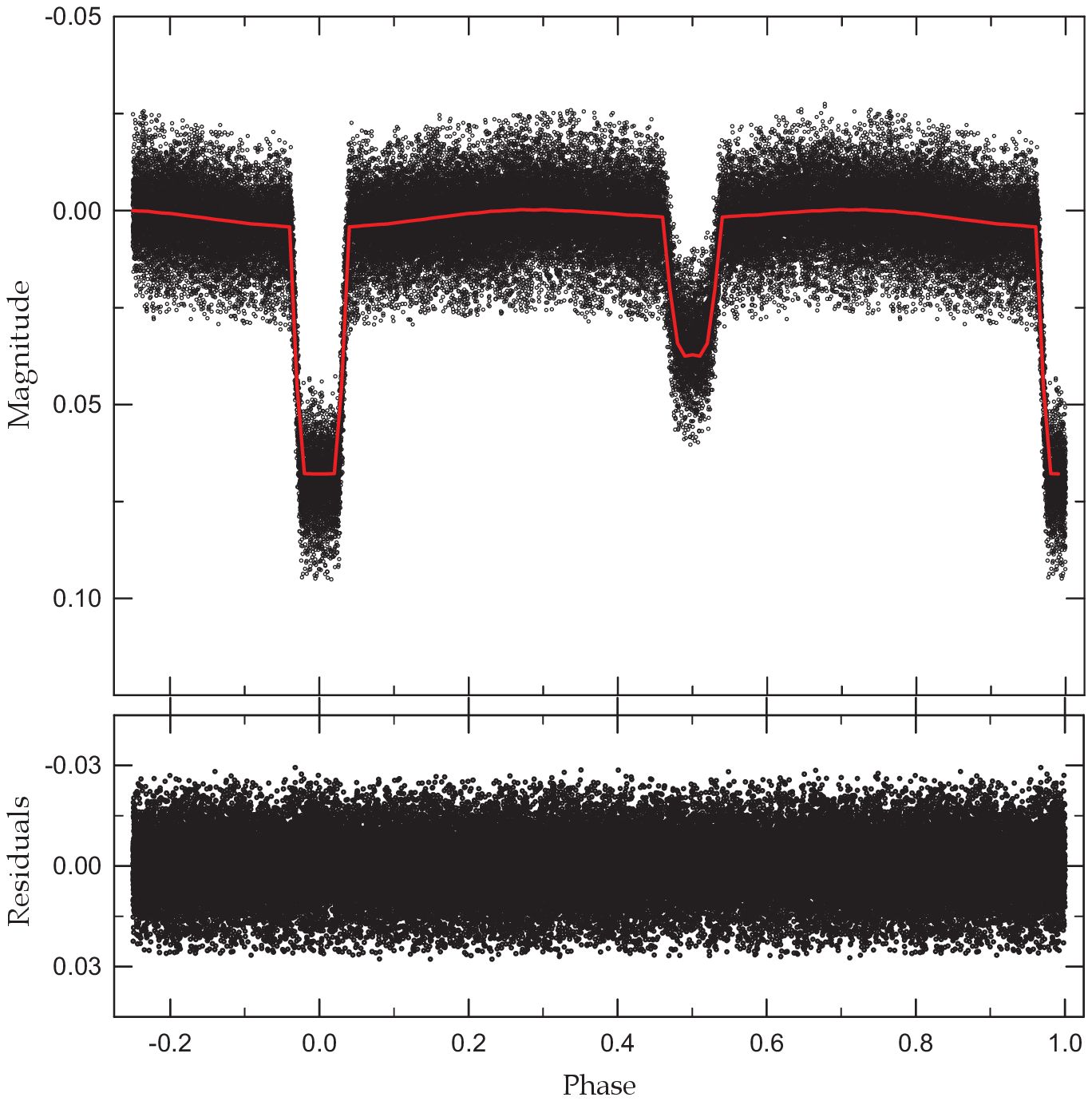}
\caption{Light curve of WASP 1625-04. The circles are individual observations taken from SuperWASP and the solid line represents 
the synthetic curve obtained with our binary parameters listed in Table 3. The corresponding residuals from this model are plotted 
in the lower panel. } 
\label{Fig1}
\end{figure}

\begin{figure}
\includegraphics[scale=0.9]{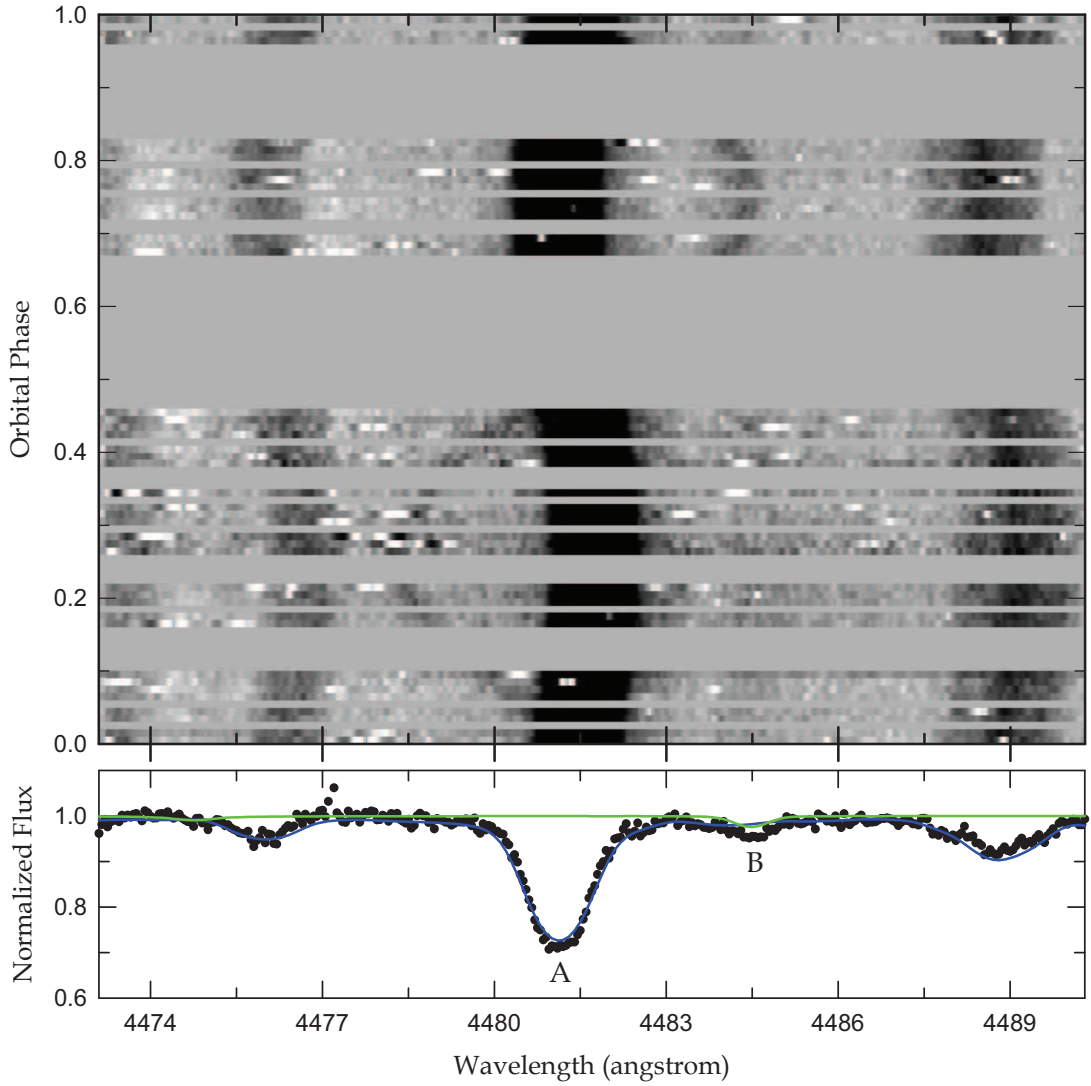}
\caption{Upper panel displays the trailed spectra of WASP 1625-04 in the Mg II $\lambda$4481 region. In the lower panel, 
the circles and blue and green lines represent the observed spectrum at an orbital phase of 0.74 (HJD 2,458,633.1240) and 
the synthetic spectra of the primary (WASP 1625-04 A; $T_{\rm eff,1}$ = 8990 K, $\log$ $g_1$ = 4.3, $v_1\sin$$i$ = 53 km s$^{-1}$) 
and secondary (WASP 1625-04 B; $T_{\rm eff,2}$ = 12,500 K, $\log$ $g_2$ = 4.8, $v_2\sin$$i$ = 10 km s$^{-1}$) components, respectively. }
\label{Fig2}
\end{figure}

\begin{figure}
\includegraphics[]{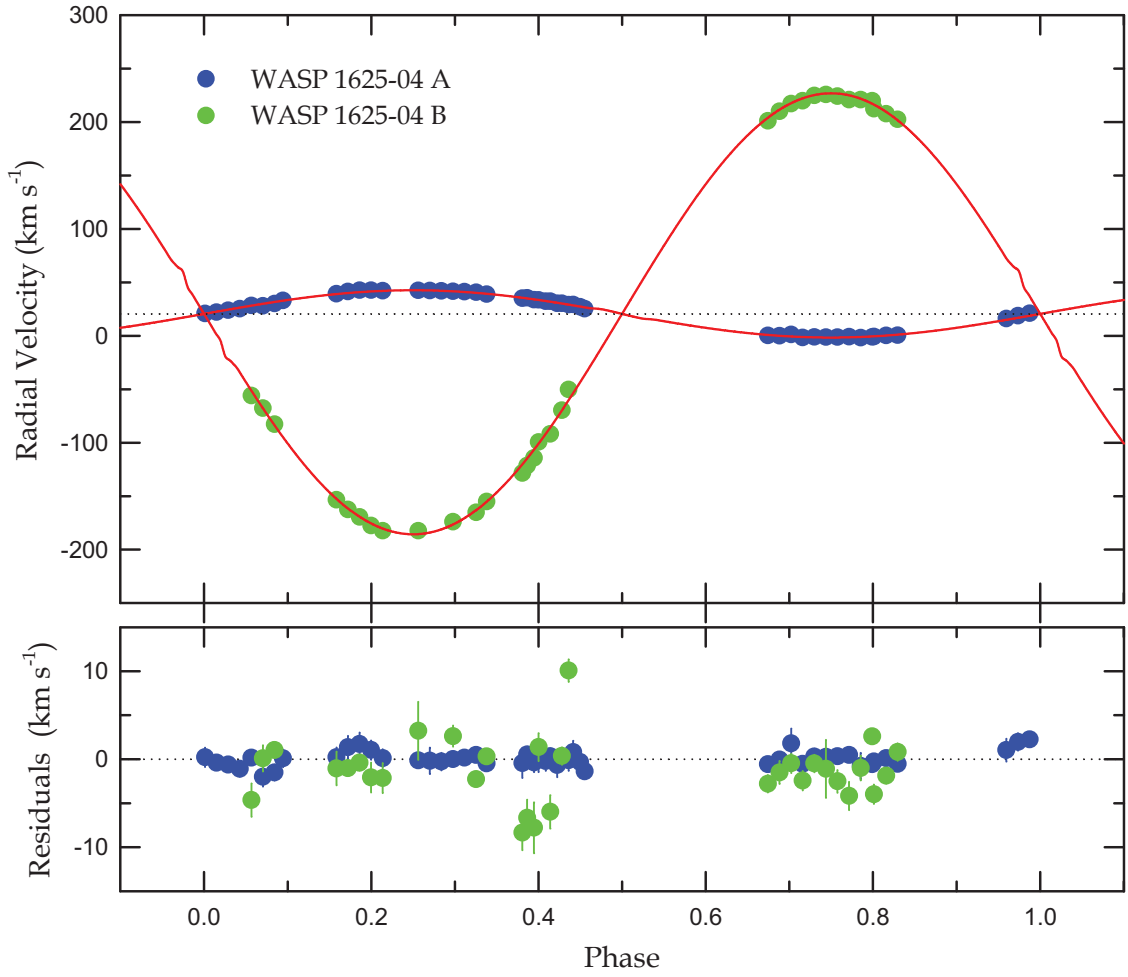}
\caption{RV curves of WASP 1625-04 with fitted models. The blue and green circles are the primary and secondary measurements, 
respectively. In the upper panel, the solid curves represent the results from a consistent light and RV curve analysis with 
the W-D program, and the dotted line denotes the system velocity of 20.54 km s$^{-1}$. The lower panel displays the residuals 
between measurements and models. }
\label{Fig3}
\end{figure}

\begin{figure}
\includegraphics{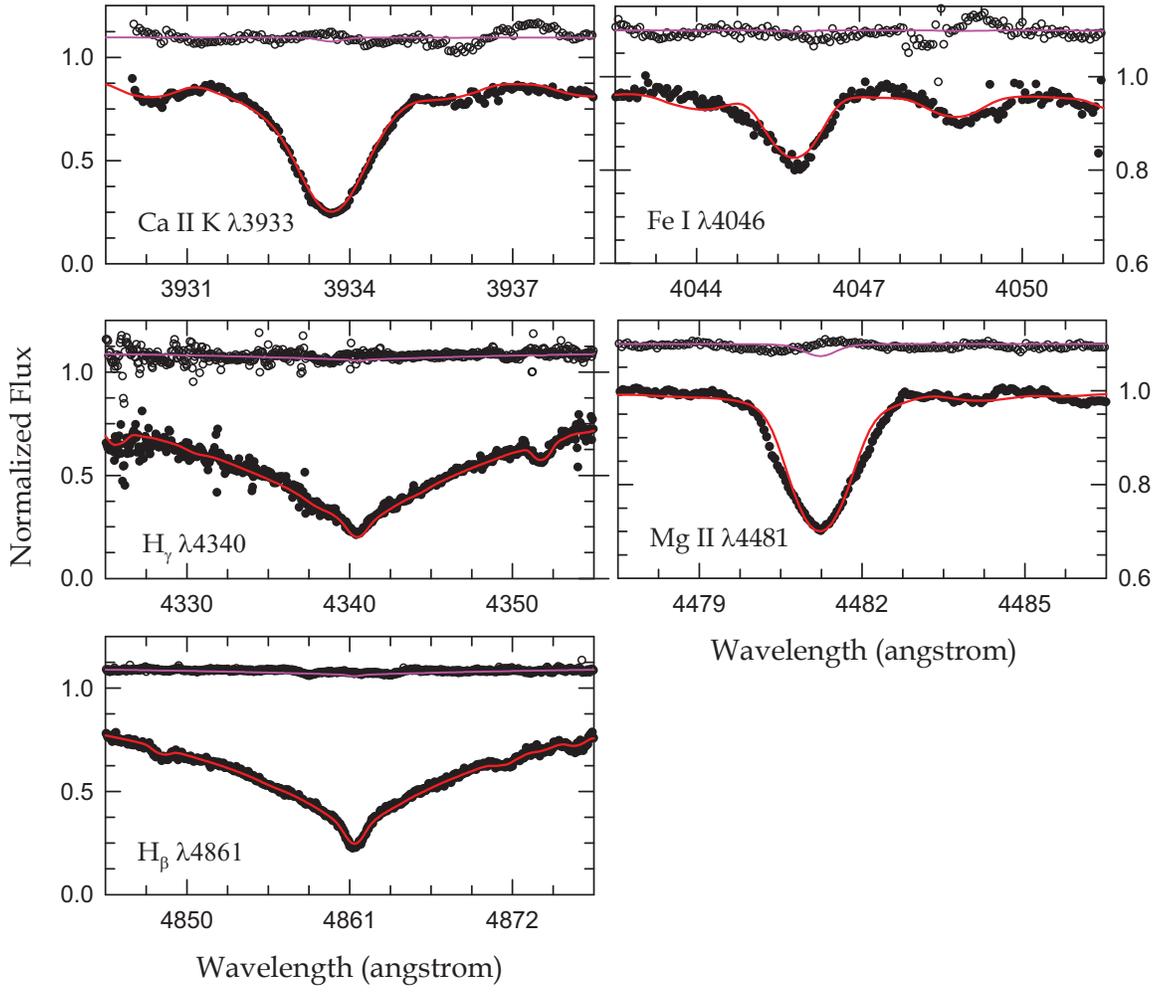}
\caption{Five spectral regions of WASP 1625-04 AB. The filled and open circles are the disentangling spectra of the primary (A) 
and secondary (B) stars, respectively, obtained with the FDB\textsc{inary} code. The red and pink lines represent their synthetic spectra 
using the same values as in Figure 2.} 
\label{Fig4}
\end{figure}

\begin{figure}
\includegraphics[scale=0.7]{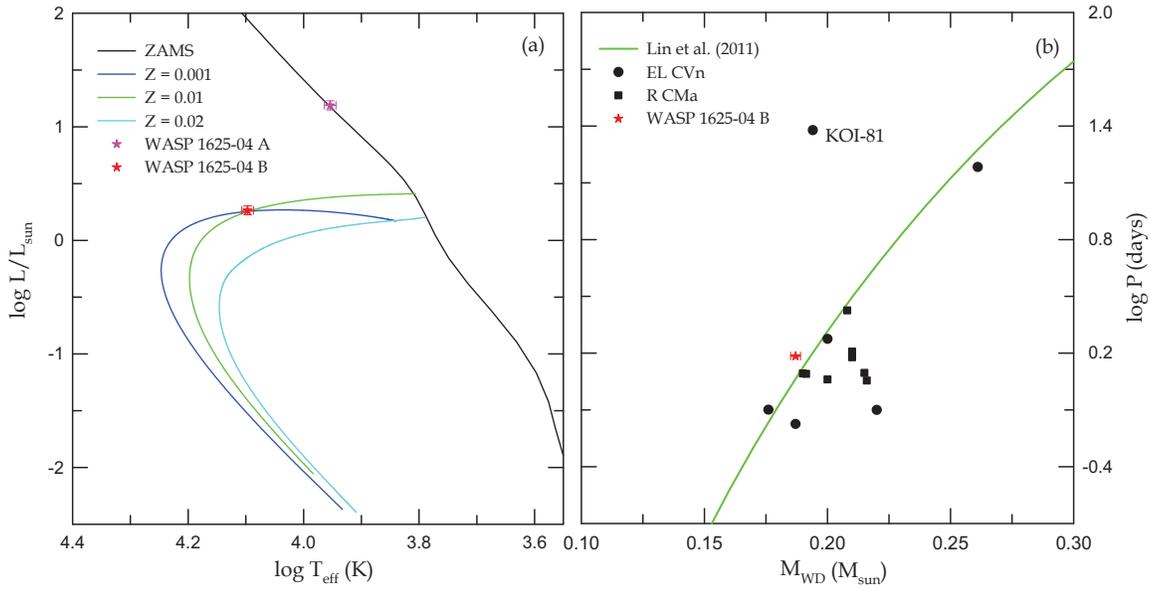}
\caption{Position of WASP 1625-04 AB (star symbols) in the HR and $M_{\rm WD}-\log P$ diagrams. In the left panel, the black line 
denotes the ZAMS stars, and the three-colored curves are the evolutionary tracks of a 0.19 M$_\odot$ He WD with different metallicities 
$Z$ from the models of Istrate et al. (2016). In the right panel, the circle and square symbols refer to EL CVn and R CMa stars, 
respectively, and the solid line denotes the theoretical relation adopted from Lin et al. (2011). }
\label{Fig5}
\end{figure}

\begin{table}
\caption{Radial Velocities of WASP 1625-04. }
\begin{tabular}{lcrcrc}
\hline
UT date            & HJD                    & $V_{1}$                 & $\sigma_1$              & $V_{2}$                 & $\sigma_2$              \\ %[1.0mm] \\[-2.0ex]
                   &                        & (km s$^{-1}$)           & (km s$^{-1}$)           & (km s$^{-1}$)           & (km s$^{-1}$)           \\
\hline                                         
2015-04-15         & 2,457,128.2569         & $ -0.7 $                & 0.6                     & $  212.4  $             &  1.1                    \\ 
2015-04-17         & 2,457,130.2308         & $ 33.0 $                & 1.0                     & $  \,     $             &  \,                     \\ 
2018-04-17         & 2,458,226.2283         & $ 39.3 $                & 1.1                     & $  -153.5 $             &  1.9                    \\ 
2018-04-17         & 2,458,226.2494         & $ 41.4 $                & 1.3                     & $  -162.6 $             &  0.7                    \\ 
2018-04-17         & 2,458,226.2705         & $ 42.6 $                & 1.3                     & $  -169.7 $             &  0.9                    \\ 
2018-04-17         & 2,458,226.2916         & $ 42.6 $                & 1.1                     & $  -177.7 $             &  1.7                    \\ 
2018-04-17         & 2,458,226.3127         & $ 42.2 $                & 1.0                     & $  -182.6 $             &  1.7                    \\ 
2019-03-13         & 2,458,556.2537         & $ 35.2 $                & 1.7                     & $  -128.5 $             &  2.0                    \\ 
2019-03-13         & 2,458,556.2748         & $ 33.8 $                & 1.1                     & $  -114.3 $             &  2.9                    \\ 
2019-03-13         & 2,458,556.2959         & $ 32.4 $                & 1.1                     & $  \,     $             &  \,                     \\ 
2019-03-13         & 2,458,556.3170         & $ 30.3 $                & 1.4                     & $  \,     $             &  \,                     \\ 
2019-03-13         & 2,458,556.3381         & $ 29.1 $                & 1.2                     & $  -50.2  $             &  1.3                    \\ 
2019-03-13         & 2,458,556.3593         & $ 27.1 $                & 1.5                     & $  \,     $             &  \,                     \\ 
2019-04-08         & 2,458,582.2096         & $ 35.6 $                & 0.9                     & $  -121.5 $             &  2.1                    \\ 
2019-04-08         & 2,458,582.2308         & $ 33.5 $                & 1.4                     & $  -99.4  $             &  1.6                    \\ 
2019-04-08         & 2,458,582.2519         & $ 32.3 $                & 0.9                     & $  -91.8  $             &  1.9                    \\ 
2019-04-08         & 2,458,582.2730         & $ 30.3 $                & 0.8                     & $  -69.7  $             &  1.0                    \\ 
2019-04-08         & 2,458,582.2941         & $ 29.3 $                & 1.3                     & $  \,     $             &  \,                     \\ 
2019-04-08         & 2,458,582.3152         & $ 25.3 $                & 0.9                     & $  \,     $             &  \,                     \\ 
2019-05-29         & 2,458,633.0180         & $ 0.3  $                & 0.6                     & $  201.2  $             &  1.0                    \\ 
2019-05-29         & 2,458,633.0394         & $ 0.0  $                & 0.7                     & $  210.1  $             &  1.3                    \\ 
2019-05-29         & 2,458,633.0605         & $ 1.2  $                & 1.7                     & $  217.2  $             &  1.1                    \\ 
2019-05-29         & 2,458,633.0816         & $ -1.6 $                & 0.6                     & $  219.8  $             &  1.1                    \\ 
2019-05-29         & 2,458,633.1028         & $ -1.1 $                & 0.9                     & $  224.8  $             &  1.0                    \\ 
2019-05-29         & 2,458,633.1240         & $ -1.3 $                & 0.7                     & $  225.7  $             &  3.3                    \\ 
2019-05-29         & 2,458,633.1451         & $ -1.2 $                & 0.5                     & $  224.2  $             &  1.3                    \\ 
2019-05-29         & 2,458,633.1663         & $ -0.9 $                & 0.7                     & $  220.9  $             &  1.6                    \\ 
2019-05-29         & 2,458,633.1874         & $ -1.8 $                & 1.5                     & $  220.9  $             &  1.4                    \\ 
2019-05-29         & 2,458,633.2086         & $ -1.1 $                & 0.9                     & $  219.8  $             &  0.7                    \\ 
2019-05-29         & 2,458,633.2337         & $ 0.4  $                & 0.7                     & $  207.8  $             &  0.9                    \\ 
2019-05-29         & 2,458,633.2548         & $ 0.6  $                & 0.8                     & $  202.6  $             &  1.0                    \\ 
2019-06-03         & 2,458,638.0322         & $ 16.0 $                & 1.3                     & $  \,     $             &  \,                     \\ 
2019-06-03         & 2,458,638.0534         & $ 18.8 $                & 1.0                     & $  \,     $             &  \,                     \\ 
2019-06-03         & 2,458,638.0746         & $ 21.0 $                & 0.7                     & $  \,     $             &  \,                     \\ 
2019-06-03         & 2,458,638.0957         & $ 20.9 $                & 1.1                     & $  \,     $             &  \,                     \\ 
2019-06-03         & 2,458,638.1168         & $ 22.2 $                & 0.5                     & $  \,     $             &  \,                     \\ 
2019-06-03         & 2,458,638.1381         & $ 23.9 $                & 0.7                     & $  \,     $             &  \,                     \\ 
2019-06-03         & 2,458,638.1592         & $ 25.3 $                & 0.9                     & $  \,     $             &  \,                     \\ 
2019-06-03         & 2,458,638.1805         & $ 28.4 $                & 0.7                     & $  -55.9  $             &  1.9                    \\ 
2019-06-03         & 2,458,638.2016         & $ 28.0 $                & 1.1                     & $  -67.7  $             &  1.5                    \\ 
2019-06-03         & 2,458,638.2228         & $ 30.2 $                & 0.9                     & $  -82.7  $             &  0.9                    \\ 
2020-04-25         & 2,458,965.1181         & $ 42.5 $                & 0.6                     & $  -182.5 $             &  3.3                    \\ 
2020-04-25         & 2,458,965.1393         & $ 42.3 $                & 1.5                     & $  \,     $             &  \,                     \\ 
2020-04-25         & 2,458,965.1604         & $ 41.9 $                & 1.0                     & $  \,     $             &  \,                     \\ 
2020-04-25         & 2,458,965.1815         & $ 41.7 $                & 0.9                     & $  -174.1 $             &  1.2                    \\ 
2020-04-25         & 2,458,965.2026         & $ 41.2 $                & 0.8                     & $  \,     $             &  \,                     \\ 
2020-04-25         & 2,458,965.2238         & $ 40.7 $                & 0.7                     & $  -165.5 $             &  0.9                    \\ 
2020-04-25         & 2,458,965.2428         & $ 38.9 $                & 0.6                     & $  -155.0 $             &  0.7                    \\ 
\hline
\end{tabular}
\end{table}

\begin{table}
\caption{Orbital Elements of WASP 1625-04 Derived with Sine-Curve Fits. }
\begin{tabular}{lcc}
\hline
Parameter                     & Primary                   & Secondary                 \\
\hline                                         
$T_0$ (HJD)                   & \multicolumn{2}{c}{2,458,638.0957$\pm$0.0019}         \\
$P$ (day)$\rm ^a$             & \multicolumn{2}{c}{1.5263234}                         \\
$\gamma$ (km s$^{-1}$)        & 20.60$\pm$0.13            & 19.30$\pm$0.66            \\
$K$ (km s$^{-1}$)             & 22.09$\pm$0.20            & 206.16$\pm$0.79           \\ %[1.0mm]
$a\sin$$i$ ($R_\odot$)        & 0.666$\pm$0.006           & 6.217$\pm$0.024           \\
$M\sin ^3$$i$ ($M_\odot$)     & 1.699$\pm$0.014           & 0.182$\pm$0.002           \\
$q$ (= $M_{\rm B}/M_{\rm A}$) & \multicolumn{2}{c}{0.1072$\pm$0.0011}                 \\
rms (km s$^{-1}$)             & 0.84                      & 3.5                       \\
\hline
\multicolumn{3}{l}{$^a$ Fixed.} \\
\end{tabular}
\end{table}

\begin{table}
\caption{Light and RV Parameters of WASP 1625-04. }
\begin{tabular}{lcc}
\hline
Parameter                         & Primary            & Secondary                    \\ 
\hline                                                                                      
$T_0$ (HJD)                       & \multicolumn{2}{c}{2,454,973.39195$\pm$0.00062}   \\
$P$ (day)                         & \multicolumn{2}{c}{1.52632329$\pm$0.00000056}     \\
$a$ (R$_\odot$)                   & \multicolumn{2}{c}{6.946$\pm$0.024}               \\
$\gamma$ (km s$^{-1}$)            & \multicolumn{2}{c}{20.54$\pm$0.12}                \\
$K_1$ (km s$^{-1}$)               & \multicolumn{2}{c}{22.16$\pm$0.19}                \\
$K_2$ (km s$^{-1}$)               & \multicolumn{2}{c}{206.41$\pm$0.73}               \\
$q$                               & \multicolumn{2}{c}{0.1074$\pm$0.0010}             \\
$i$ (deg)                         & \multicolumn{2}{c}{82.94$\pm$0.14}                \\
$T$ (K)                           & 8990$\pm$200       & 12,500$\pm$270               \\
$\Omega$                          & 4.398$\pm$0.016    & 4.013$\pm$0.018              \\
$\Omega_{\rm in}$$\rm ^a$         & \multicolumn{2}{c}{1.982}                         \\
$X$, $Y$                          & 0.655, 0.105       & 0.741, 0.076                 \\
$x$, $y$                          & 0.549, 0.250       & 0.451, 0.233                 \\
$l/(l_1+l_2)$                     & 0.9437$\pm$0.0004  & 0.0563$\pm$0.0004            \\
$r$ (pole)                        & 0.2329$\pm$0.0009  & 0.0418$\pm$0.0004            \\
$r$ (point)                       & 0.2352$\pm$0.0009  & 0.0419$\pm$0.0004            \\
$r$ (side)                        & 0.2346$\pm$0.0009  & 0.0418$\pm$0.0004            \\
$r$ (back)                        & 0.2350$\pm$0.0009  & 0.0419$\pm$0.0004            \\
$r$ (volume)$\rm ^b$              & 0.2342$\pm$0.0009  & 0.0418$\pm$0.0004            \\ 
\hline
\multicolumn{3}{l}{$^a$ Potential for the inner critical Roche surface.} \\
\multicolumn{3}{l}{$^b$ Mean volume radius.} 
\end{tabular}
\end{table}

\begin{table}
\caption{Absolute Parameters of WASP 1625-04. }
\begin{tabular}{lccc}
\hline
Parameter                     & Primary             & Secondary                   \\                                                                                         
\hline 
$M$ ($M_\odot$)               & 1.745$\pm$0.013     & 0.187$\pm$0.002             \\
$R$ ($R_\odot$)               & 1.626$\pm$0.008     & 0.290$\pm$0.003             \\
$\log$ $g$ (cgs)              & 4.258$\pm$0.005     & 4.785$\pm$0.010             \\
$\rho$ ($\rho_\odot$)         & 0.407$\pm$0.007     & 7.68$\pm$0.25               \\
$v_{\rm sync}$ (km s$^{-1}$)  & 53.91$\pm$0.27      & 9.62$\pm$0.10               \\
$v\sin$$i$ (km s$^{-1}$)      & 53$\pm$5            & \,                          \\
$T_{\rm eff}$ (K)             & 8990$\pm$200        & 12,500$\pm$270              \\
$L$ ($L_\odot$)               & 15.5$\pm$1.4        & 1.84$\pm$0.16               \\
$M_{\rm bol}$ (mag)           & 1.76$\pm$0.10       & 4.07$\pm$0.10               \\
BC (mag)                      & $-$0.06$\pm$0.03    & $-$0.78$\pm$0.05            \\
$M_{\rm V}$ (mag)             & 1.82$\pm$0.10       & 4.85$\pm$0.11               \\
Distance (pc)                 & \multicolumn{2}{c}{420$\pm$22}                    \\
\hline
\end{tabular}
\end{table}

\bsp
\label{lastpage}
\end{document}